\documentclass[journal]{IEEEtran}
\usepackage{cite}
\usepackage{subfig}
\usepackage{amsmath,amssymb,amsfonts,amsthm}
\usepackage{algorithmic}
\usepackage{graphicx}
\usepackage{hyperref}
\usepackage{multirow}
\usepackage{multicol}
\usepackage{booktabs}
\hypersetup{colorlinks=true,urlcolor=red}
\usepackage{bm} 
\usepackage{xspace}

\long\def\comment#1{} 

\newcommand{\xmath}[1] {\ensuremath{#1}\xspace}
\newcommand{\blmath}[1] {\xmath{\bm{#1}}}

\newcommand{\1}{\blmath{1}}

\newcommand{\Xb}{{\blmath X}}
\newcommand{\Yb}{{\blmath Y}}

\newcommand{\xb}{{\blmath x}}
\newcommand{\yb}{{\blmath y}}
\newcommand{\zb}{{\blmath z}}

\newcommand{\Tc}{\mathcal{T}}
\newcommand{\Xc}{\mathcal{X}}
\newcommand{\Yc}{\mathcal{Y}}

\newcommand{\Rd}{{\mathbb R}}

\newcommand{\beq}{\begin{equation}}
\newcommand{\eeq}{\end{equation}}
\newcommand{\beqa}{\begin{eqnarray}}
\newcommand{\eeqa}{\end{eqnarray}}

\usepackage{array}
\usepackage{multirow}

\usepackage{cite}
\usepackage{amsmath}
\usepackage{amsfonts}

\begin{document}

\title{AdaIN-Switchable CycleGAN for Efficient Unsupervised Low-Dose CT Denoising }

\author{Jawook~Gu,
        and~Jong~Chul~Ye,~\IEEEmembership{Fellow,~IEEE}
\thanks{J. Gu and J. C. Ye are with the Department
of Bio and Brain Engineering, Korea Advanced Institute of Science and Technology, Daejeon 305-701,
Korea, e-mail: (jwisdom9299@kaist.ac.kr; jong.ye@kaist.ac.kr).}
\thanks{}}

\maketitle

\begin{abstract}
Recently, deep learning approaches have been extensively studied for low-dose CT denoising
thanks to its superior performance despite the fast computational time.
In particular, cycleGAN has been demonstrated as a powerful unsupervised learning  scheme to improve the low-dose CT image quality without requiring matched high-dose reference data.
Unfortunately, one of the main limitations of the cycleGAN approach is that it requires two deep neural network generators at the training phase, although only one of them is used at the inference phase.
The secondary auxiliary generator is needed to enforce the cycle-consistency, but the additional memory requirement and increases of the learnable parameters are the main huddles for cycleGAN training.
To address this issue, here we propose a novel cycleGAN architecture using a single {\em switchable} generator.
In particular, a single generator is implemented using adaptive instance normalization (AdaIN) layers 
so that the baseline generator converting a low-dose CT image to a routine-dose CT image 
can be switched to a generator converting high-dose to low-dose by simply changing the AdaIN code.
Thanks to the shared baseline network, the additional memory requirement and weight increases are minimized,
and the training can be done more stably even with small training data.
Experimental results show that the proposed method outperforms the previous cycleGAN approaches
while using only about half the parameters.
\end{abstract}

\begin{IEEEkeywords}
Low-dose CT, deep learning, unsupervised learning, cycleGAN, adaptive instance normalization (AdaIN), wavelet transform
\end{IEEEkeywords}

\IEEEpeerreviewmaketitle

\section{Introduction}

\IEEEPARstart{X}{-RAY} computed tomography (CT) is one of the most widely used medical imaging modalities 
thanks to the advantage of being able to obtain high-quality medical images quickly.
Unfortunately, CT also has disadvantages in that radiation passing through a patient can increase the risk of cancer. 
For this reason, low-dose CT, which acquires images with only a small amount of radiation, is used in many cases. 
However, a low-dose CT image has a high noise level and low image contrast compared to a CT image taken with a full dose, which makes it difficult to accurately diagnose.

\begin{figure*}[!t]
\centering
\includegraphics[width=0.9\textwidth]{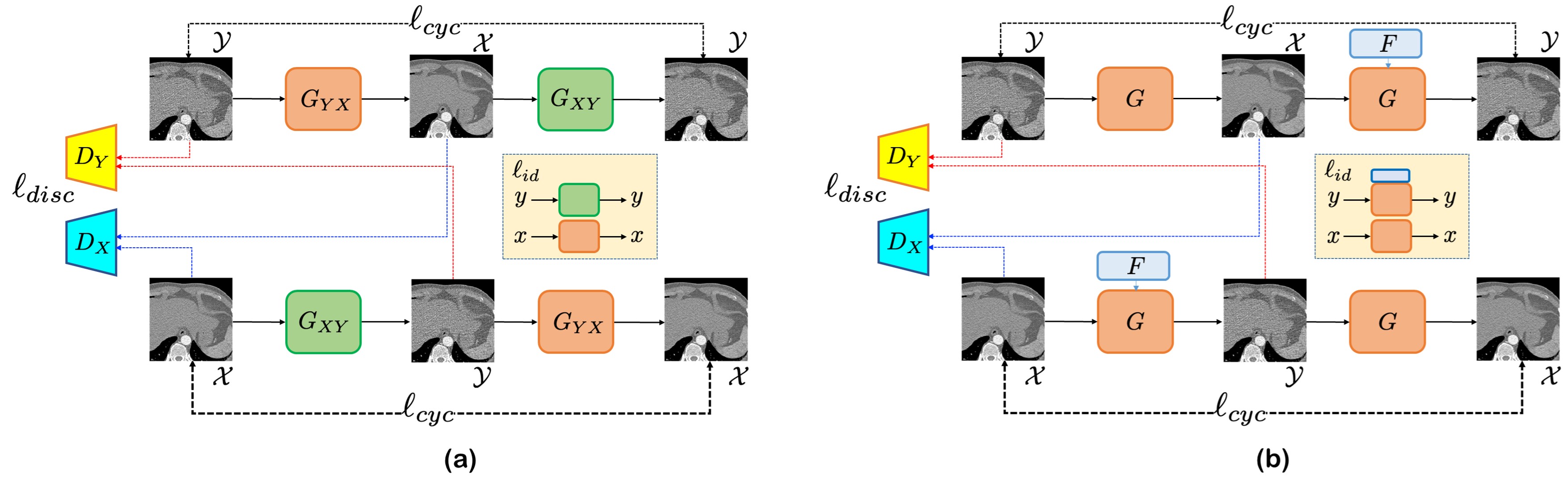}
\caption{Overview of unsupervised learning methods. \(\Xc\) and \(\Yc\) denote high-dose CT domain and low-dose CT domain, respectively. The networks \(D_X\) and \(D_Y\) discriminate fake images and real images in $\Xc$ and $\Yc$, respectively. (a) CycleGAN with two different generators, \(G_{XY}\) and \(G_{YX}\). (b) AdaIN-switchable cycleGAN. The generator $G$  taking the vector from AdaIN-code generator $F$ converts high-dose images to low-dose images. All the networks are trained using the adversarial loss \(\ell_{disc}\), cycle loss \(\ell_{cyc}\) and identity loss \(\ell_{id}\).}
\label{fig_cycleGAN}
\end{figure*}

Various studies have been conducted to improve the image quality of low-dose CT by removing noises. 
Low-dose CT denoising algorithms traditionally used the model-based iterative reconstruction (MBIR) methods \cite{beister2012iterative, leipsic2010adaptive, renker2011evaluation}. 
An MBIR approach repeatedly performs forward projection and back-projection, which requires excessive amount of computations.

Deep learning-based image processing algorithms have shown remarkable achievements in the computer vision field over the recent years
\cite{krizhevsky2012imagenet,ronneberger2015u,zhang2016beyond,kim2015accurate,shi2016real}.
There are two major advantages of deep learning:
(1) it shows robust performance for various inputs without adjusting hyperparameters because the network learns image features directly from the data, and 
(2) image processing is performed quickly without iterative computations.
Thanks to these advantages, in the CT imaging field, many studies using deep learning algorithm have been conducted. 
Kang et al. \cite{kang2017deep, kang2018deep}, Chen et al. \cite{chen2017low}, and Yang et al. \cite{yang2018low} have published studies using deep learning for low-dose CT denoising. 
These early stage studies using deep learning were mostly conducted by supervised learning methods that require training data composed of pairs of low-dose  and matched high-dose CT images. 
However, in a clinical environment, it is difficult to obtain matched low- and high- dose data due to the increase of the total radiation dose to patients in
the acquisition of the paired data.
Moreover, the simulation-based addition of the noises often produces biased results that are not matched to the real acquisition scenarios.
Therefore, unsupervised learning, which can train a convolutional neural network (CNN) without a paired dataset, has attracted much attention.

Generative adversarial network (GAN) \cite{goodfellow2014generative} is a popular unsupervised learning framework.
CycleGAN is one of the most successful image-to-image translation algorithm between an input-image domain and a target-image domain without a paired dataset \cite{zhu2017unpaired}. 
Inspired by this, our group has also published an unsupervised low-dose CT denoising study using cycleGAN \cite{kang2019cycle}.
Furthermore, our follow-up study \cite{sim2019optimal} has revealed the mathematical origin of the cycleGAN from optimal transport theory \cite{peyre2019computational,villani2008optimal},
showing that cycleGAN can be derived as a variation formulation that minimizes the statistical distances between empirical distributions and ``push-forward'' distribution in the measurement and the image domain simultaneously \cite{sim2019optimal}. 

Despite its superior performance with strong theoretical foundations,
CycleGAN is not free of limitations.  
In particular, cycleGAN requires two separate generators to enforce the cycle-consistency by performing image-to-image translation in the forward and opposite directions simultaneously.
For example, in the denoising task, one generator performs the noise reduction, and the other performs the noise generation during network training.
After the training is completed, only the generator that performs noise reduction is used.
Accordingly, the additional GPU memory requirement by the auxiliary generator is inefficient and prevents the use of deeper neural network for the noise reduction generator.
Moreover, increased number of trainable weights in neural networks due to the second generator often requires a large amount of training dataset.

To address this fundamental issue,  here we propose a novel cycleGAN architecture using a single generator with adaptive instance normalization (AdaIN) layers.
AdaIN was originally proposed as an image style transfer method, 
in which the mean and variance of the feature vectors are replaced by those of the style reference image \cite{huang2017arbitrary}.
Recently, AdaIN has attracted attention as a powerful component to control generative models.
For example, in the StyleGAN \cite{karras2019style}, the network generated various realistic faces with the same inputs by simply adjusting the mean and variance of the feature maps using AdaIN.
Furthermore, recent study \cite{mroueh2019wasserstein} has shown that the change of style by AdaIN is not a cosmetic change but a real one thanks to the important link to the optimal transport \cite{peyre2019computational,villani2008optimal} between the feature vector spaces.

Inspired by the success of AdaIN and its theoretical understanding, 
one of the most important contributions of this work is to demonstrate that
a {\em single} generator with AdaIN layers can learn the bidirectional image-to-image translation function which is essential for the learning using the cycle-consistent loss.
Thus, the network can be trained in the similar way of cycleGAN without an additional generator, which is not actually necessary for the inference.
This results in many advantages which may address the limitations of the existing cycleGAN.
First, the AdaIN code generation can be easily done with a very light weight network, dubbed {\em AdaIN code generator}, 
so the memory requirement for the generator of our method is only about a half of the original one.
Furthermore, once the neural network is trained, the AdaIN code is no more necessary, 
and the simple baseline network is used as a denoising network, 
which makes the system very simple.
Finally, due to the same underlying network structure, the number of trainable weights is much smaller so that training can be more easier than the original cycleGAN counterpart. 
In fact, our experimental results demonstrated that our network can be trained with much smaller training dataset, which was not possible using the existing cycleGAN.

To further improve the performance, our method is combined with the wavelet residual proccessing.
More specifically, Song et al. \cite{song2020unsupervised} have shown that the preprocessing method using wavelet transform is effective in separating the common components of an image from noises, so that the CNN can learn the noise better. 
Accordingly, the network trained using wavelet high-frequency detail images can improve noise reduction performance and preserve the information in the original CT image.
Taken together, this study introduces a novel AdaIN-switchable cycleGAN for low-dose CT denoising using wavelet residual signals.
Experimental results confirmed that the proposed method has a higher memory efficiency and higher noise reduction performance compared to the existing cycleGAN.

\begin{figure*}[t!]
\centering
\includegraphics[width=1.0\textwidth]{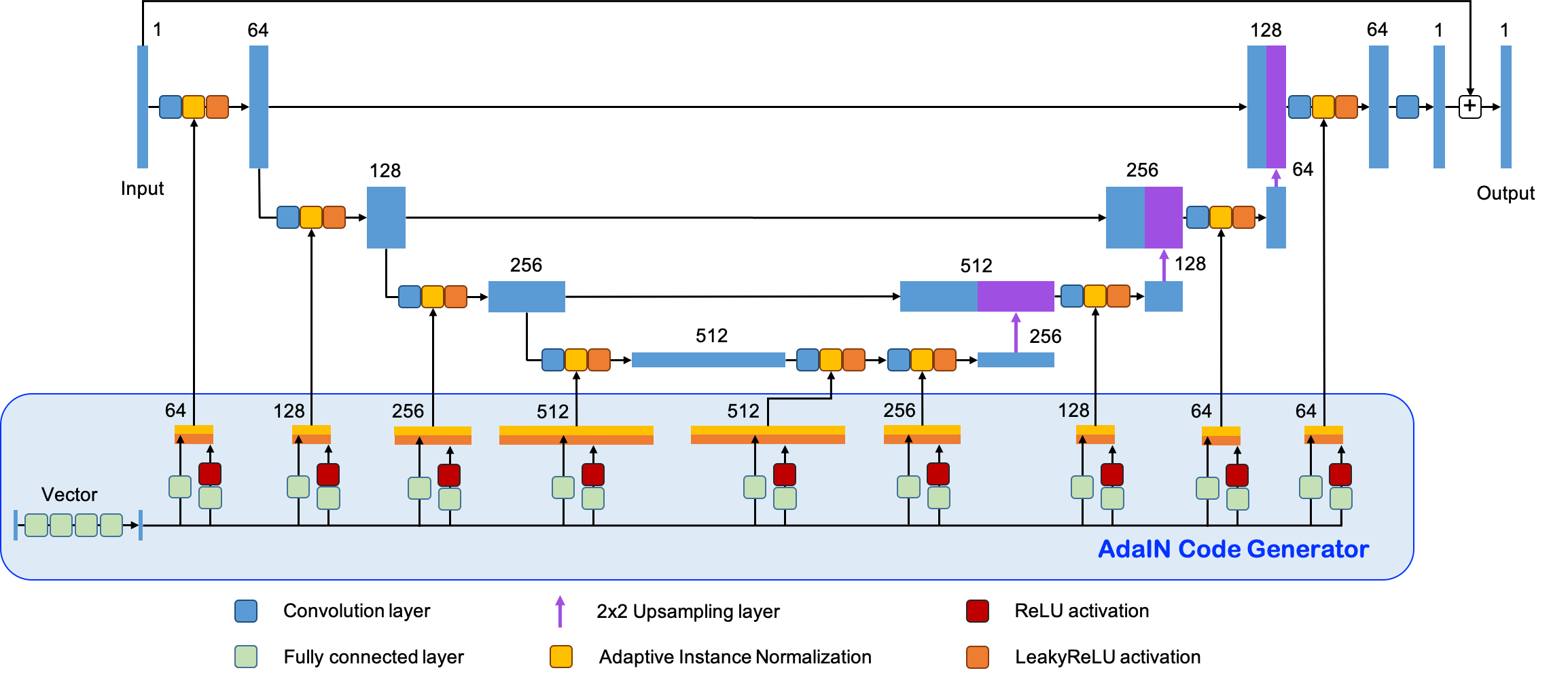}
\caption{Architecture of image generator connected to the AdaIN code generator. 
The image generator has a structure of U-Net. 
The AdaIN code generator consists of fully connected layers and ReLU activation layers. 
Due to the small size of input and output, 
the AdaIN code generator is much lighter compared to the image generator. 
The outputs of the AdaIN code generator are connected to the AdaIN layers in the image generator.}
\label{fig_network}
\end{figure*}


\section{Related works}

\subsection{Deep Learning Approaches for Low-Dose CT Denoising}

Many deep learning approaches for low-dose CT denoising have been published for recent years.
Kang et al. \cite{kang2017deep} have first proposed a low-dose CT denoising network, called AAPM-Net, in the 2016 AAPM Low-dose CT Grand Challenge.
Chen et al. have introduced the low-dose CT denosing algorithm using residual encoder-decoder convolution neural network (RED-CNN) \cite{chen2017low}.
Yang et al. used Wasserstein generative adversarial network for low-dose CT denosing \cite{yang2018low}.
These early works were based on the supervised learning framework. 
For example, the dataset from the 2016 AAPM low-dose CT Grand Challenge consists of the matched high-dose CT images and simulated low-dose CT images and was used for supervised learning in these works.

To train a neural network without matched reference data, Wolterink et al. \cite{wolterink2017generative} have proposed low-dose CT denoising network using an adversarial loss.
To prevent from generating spurious features from the adversarial loss, 
Kang et al. \cite{kang2019cycle} have employed cycleGAN for low-dose CT denoising by imposing the cycle-consistency loss.
The results showed that the cycle-consistency and identity losses played an important role to preserve the detail information of the original CT image without generating spurious features \cite{kang2019cycle}.

\subsection{Deep Learning using Wavelet High Frequency Images}

Wavelet transform is known for its capability of decomposing images into directional components \cite{zhou2005nonsubsampled}.
In fact, the AAPM-Net by Kang et al. \cite{kang2017deep} has applied contourlet transform preprocessing to the CT images, 
so that deep learning can be performed in contourlet transform domain to learn the directional components of noises.

Yet another advantage of using a wavelet transform for deep learning is that 
the directional components to be modified can be separated from the other bands based on the prior knowledge,
so that the network can easily focus on learning the noise components.
For example, in a recent paper by Song et al. \cite{song2020unsupervised},
only a subset of the wavelet bands was used to recompose a wavelet directional image, 
and the neural network was designed to learn the mapping between these directional images in an unsupervised manner using cycleGAN.

\subsection{Adaptive Instance Normalization (AdaIN)}

AdaIN \cite{huang2017arbitrary} is an extension of instance normalization \cite{ulyanov2016instance}, 
which goes beyond the classical role of the normalization methods for image style transfer.
The key idea of AdaIN is that latent space exploration is possible 
by adjusting the mean and variance of the feature map, 
so that the outputs with different styles can be generated with the same input image.
Specifically, style transfer was performed by matching the mean and variance of the feature map of the input image to those of the style reference image \cite{huang2017arbitrary}.

More specifically, let a multi-channel feature map be represented by
\begin{align}\label{eq:X}
\Xb =&\begin{bmatrix} \xb_1  & \cdots &\xb_C \end{bmatrix} \in \Rd^{HW\times C},
\end{align}
where 
$\xb_n  \in \Rd^{HW\times 1}$ refers to the $n$-th column vector of $\Xb$, 
which represents the vectorized feature map of size of $H\times W$  at the $n$-th channel.
Suppose, furthermore, the corresponding feature map for the style reference image is given by
 \begin{align}\label{eq:Y}
\Yb =&\begin{bmatrix} \yb_1  & \cdots &\yb_C \end{bmatrix} \in \Rd^{HW\times C},
\end{align}
Then, AdaIN changes the feature data for each channel using the following transform:
\begin{align}
\zb_n &= \Tc(\xb_n,\yb_n),\quad n=1,\cdots, C
\end{align}
where
\begin{align}\label{eq:AdaIN}
\Tc(\xb,\yb)  := \frac{\sigma(\yb) }{\sigma(\xb) }\left(\xb -\mu(\xb) \1\right) +\mu(\yb)\1,\quad 
\end{align}
where $\1 \in \Rd^{HW}$ is the $HW$-dimensional vector composed of 1, and 
 $\mu(\xb)$ and $\sigma(\xb)$ are channel-wise mean and standard deviation (std) for $\xb\in \Rd^{HW}$:
\begin{align}
\mu(\xb)  =  \frac{1}{HW}\1^\top \xb ,~&
\sigma(\xb)  = \sqrt{ \frac{1}{HW} \| \xb  -\mu \1\|^2}
\end{align}
Note that the instance normalization \cite{ulyanov2016instance} only differs the choice of the $\sigma(\yb)$ and $\mu(\yb)$.
Specifically, rather than estimating the $\sigma(\yb)$ and $\mu(\yb)$ using the style reference image, the instance normalization estimates them from the input features using backpropagation.

At glance, the AdaIN transform in \eqref{eq:AdaIN} appears heuristic.
However, the in-depth study \cite{mroueh2019wasserstein} has shown that
the AdaIN transform in \eqref{eq:AdaIN} is the optimal transport scheme between two probability
spaces $\Xc \subset \Rd^{HW}$ and $\Yc \subset \Rd^{HW}$, 
which are equipped with the i.i.d. Gaussian distributions with (mean,std)= $(\mu(\xb),\sigma(\xb))$  and
$(\mu(\yb),\sigma(\yb))$, respectively.
Therefore, the AdaIN transform provides the minimal cost transport path when converting one feature map to another.

\section{Theory}

\subsection{Overall Scheme}

Since the low-dose CT noises are mainly in the high frequency range,
the high-frequency and low-frequency components of the CT image are decomposed 
using wavelet transform in our task, 
so that the neural network is trained using the high-frequency component only 
while maintaining the low-frequency component.
Furthermore, to make the neural network training memory and data efficient,
two opposite image-to-image translation tasks, noise reduction and noise generation, 
are conducted by a single network with AdaIN.
The detailed description of each component is as follows.

\subsection{Switchable Generator using AdaIN Layers}

As shown in Fig.~\ref{fig_cycleGAN}(a),  
the cycleGAN framework for low-dose CT denoising requires two generators: 
one from low-dose to high-dose images $(G_{YX})$, 
the other from high-dose to low-dose images ($G_{XY}$).
Inspired by the observation that AdaIN is an optimal transport plan between the feature layers,
our goal is to implement two generators using a single baseline network followed by optimal transport layers to specific target distributions.
More specifically, for the low-dose to high-dose generator, 
we use the autoencoder as the baseline network
and then use AdaIN transform to transport the autoencoder features to the high-dose features. Similarly, for the high-dose to low-dose generator, 
the autoencoder features are then transported to the low-dose features using another AdaIN transform.
Furthermore, a key simplification comes from that if the first AdaIN code for the low-dose to high-dose generator is set to a constant vector, 
then only the second AdaIN code needs to be optimized through the additional neural network $F$ as shown in Fig.~\ref{fig_cycleGAN}(b).

More specifically, the architecture of the proposed generator network is shown in Fig. \ref{fig_network}.
Basic structure of the autoencoder network is from U-Net \cite{ronneberger2015u}.
In the proposed network, the batch normalization layers are replaced with AdaIN layers.
The AdaIN layers take a mean vector and a variance vector as input.
During the noise reduction task, the mean and variance vectors for the AdaIN layers are constant.
Value of the mean vector is set to zero, and value of the variance vector is set to one.
This gives us an advantage of not using the AdaIN code generator at the inference phase.
During the noise generation task, the AdaIN layers take the output vectors from the AdaIN code generator which consists of fully connected layers and is trained together with the cycleGAN architecture.
Then, at the inference phase, we only use the U-Net with zero mean and unit variance vectors as the low-dose to high-dose denoising network.
Since the baseline U-Net is shared by the two generators and only the additional network generating AdaIN code is needed,
the total memory requirement can be significantly reduced as long as the AdaIN code generator is light.

In fact, the AdaIN code generator takes the ones vector with \(1 \times 128 \) size as input, and outputs nine pairs of mean and variance vectors as shown in Fig. \ref{fig_network}.
The four fully connected layers are shared parameters, and the last layers are not shared 
because the length of the output vectors must be equal to the number of channels of the feature map to normalize. 
ReLU activation is applied to prevent the variance vectors from becoming negative.
Accordingly, the AdaIN code generator is very light, whose network complexity is negligible compared to the additional generator in the conventional cycleGAN.
Thanks to the reduction of the number of network weights, 
our network can be trained robustly with a smaller training dataset,
as will be shown later in discussion.

\subsection{Cycle-Consistent Adversarial Training}

We trained the low-dose CT denoising networks in a manner similar to the cycleGAN method \cite{zhu2017unpaired}, 
since the cycleGAN is shown as an optimal transport approach 
for unsupervised learning between two probability distributions \cite{sim2019optimal}.
The difference between the proposed method and the cycleGAN is that the cycleGAN requires two generators, whereas the proposed method requires only one generator.

Figure \ref{fig_cycleGAN}(b) visualizes the learning scheme of the proposed method.
Here, let $\Xc$ and $\Yc$ be the high-dose and low-dose CT image domains, and
\(P_{\Xc}\) and \(P_{\Yc}\) be the associated probability distributions, respectively.
Unlike the cycleGAN method with two generators, the proposed method has only one generator \(G\) with AdaIN layers which are connected to the AdaIN code generator \(F\) or default AdaIN vector $c_x$ composed of zero mean and unit variance.
More specifically, with the constant code $c_x$, the generator
$G(y;c_x):\Yc\mapsto \Xc$ ``push-forwards'' the probability measure \(P_{\Yc}\) to \(P_{\Xc}\) \cite{peyre2019computational,villani2008optimal}. 
On the other hand, with the learned vector $c_y$, the generator $G(x;c_y):\Xc\mapsto \Yc$ performs the push-forward of the measure   \(P_{\Xc}\) to  \(P_{\Yc}\).
Here, the AdaIN code \(c_y\) is obtained as the output of the AdaIN code generator \(F\), where the input vector to \(F\) is ones vector \(c\) with size of \(1\times128\), 
and \(c_y = F(c)\).
Note that the mean and variance vectors for AdaIN codes must be the same length as the number of channels of the feature map.
Then, the generator learns not only the image translation function from \(P_{\Xc}\) to \(P_{\Yc}\) but also the function from \(P_{\Yc}\) to \(P_{\Xc}\) by simply changing the AdaIN code.

The training of our cycleGAN can be done by solving the following min-max problem:
\begin{align}\label{eq:minmax}
\min\limits_{G,F}\max\limits_{D_X,D_Y}\ell_{total}(G, F, D_X, D_Y)
\end{align}
where our cycleGAN loss is defined as follows:
\begin{equation} \label{eq:loss_total}
\begin{split}
\ell_{total}(G, F, D_X, D_Y) = -&\ell_{disc}(G, F, D_X, D_Y) + \\
&\lambda_{cyc} \ell_{cycle}(G, F) + \\
&\lambda_{id} \ell_{identity}(G, F)
\end{split}
\end{equation}
where $\lambda_{cyc}$ and $\lambda_{id}$ denote the weighting parameters for the cycle loss and the identity loss terms.
Here, the discriminator loss $\ell_{disc}(G, F, D_X, D_Y)$ is composed of LSGAN losses \cite{mao2017least}:
\begin{equation} \label{eq:loss_gan_x}
\begin{split}
\ell_{disc}(G, F, D_X,D_Y) = &\mathbb{E}_{y \sim P_{\mathcal{Y}}} [\|D_Y(y)\|_1] \\
& + \mathbb{E}_{x \sim P_{\mathcal{X}}} [\| 1 - D_Y(G(x; F(c)))\|_1] \\
&+\mathbb{E}_{x \sim P_{\mathcal{X}}} [\|D_X(x)\|_1] \\
& + \mathbb{E}_{y \sim P_{\mathcal{Y}}} [\| 1 - D_X(G(y; c_x))\|_1]
\end{split}
\end{equation}
where $\|\cdot\|$ is the $l_1$ norm, and $D_X$ (resp. $D_Y$) is the discriminator 
that tells the fake high-dose (resp. low-dose) images from real high-dose (resp. low-dose) images.
The cycle loss $\ell_{cyc}(G, F)$ in \eqref{eq:loss_total} is defined as: 
\begin{equation} \label{eq:loss_cycle}
\begin{split}
\ell_{cyc}(G, F) = &\mathbb{E}_{y \sim P_{\mathcal{Y}}} [\|G(G(y; c_x); F(c)) - y\|_1] \\
& + \mathbb{E}_{x \sim P_{\mathcal{X}}} [\|G(G(x; F(c)); c_x) - x\|_1]
\end{split}
\end{equation}
In addition, the identity loss in \eqref{eq:loss_total} should be designed to prevent 
the target domain image from being distorted more than necessary by GAN loss, 
so that this imposes the fixed point constraint for the algorithm.
Specifically, the images from \(P_{\Xc}\) with the AdaIN code \(c_x\) 
and the images from \(P_{\Yc}\) with the AdaIN code \(F(c)\) should not be changed by the generator, which leads to the following identity loss:
\begin{equation} \label{eq:loss_identity}
\begin{split}
\ell_{id}(G, F) = &\mathbb{E}_{y \sim P_{\mathcal{Y}}} [\|G(y; F(c)) - y\|_1] \\
& + \mathbb{E}_{x \sim P_{\mathcal{X}}} [\|G(x; c_x) - x\|_1]
\end{split}
\end{equation}

The discriminators are trained to minimize the discriminator loss \(\ell_{disc}\) 
while the generator is trained to maximize it.
The generator and discriminators are updated alternatively for adversarial training.
In the meanwhile, the AdaIN code generator $F$ is trained such that overall cycleGAN training
can be done stably.

\begin{figure}[!t]
\centering
\includegraphics[width=\columnwidth]{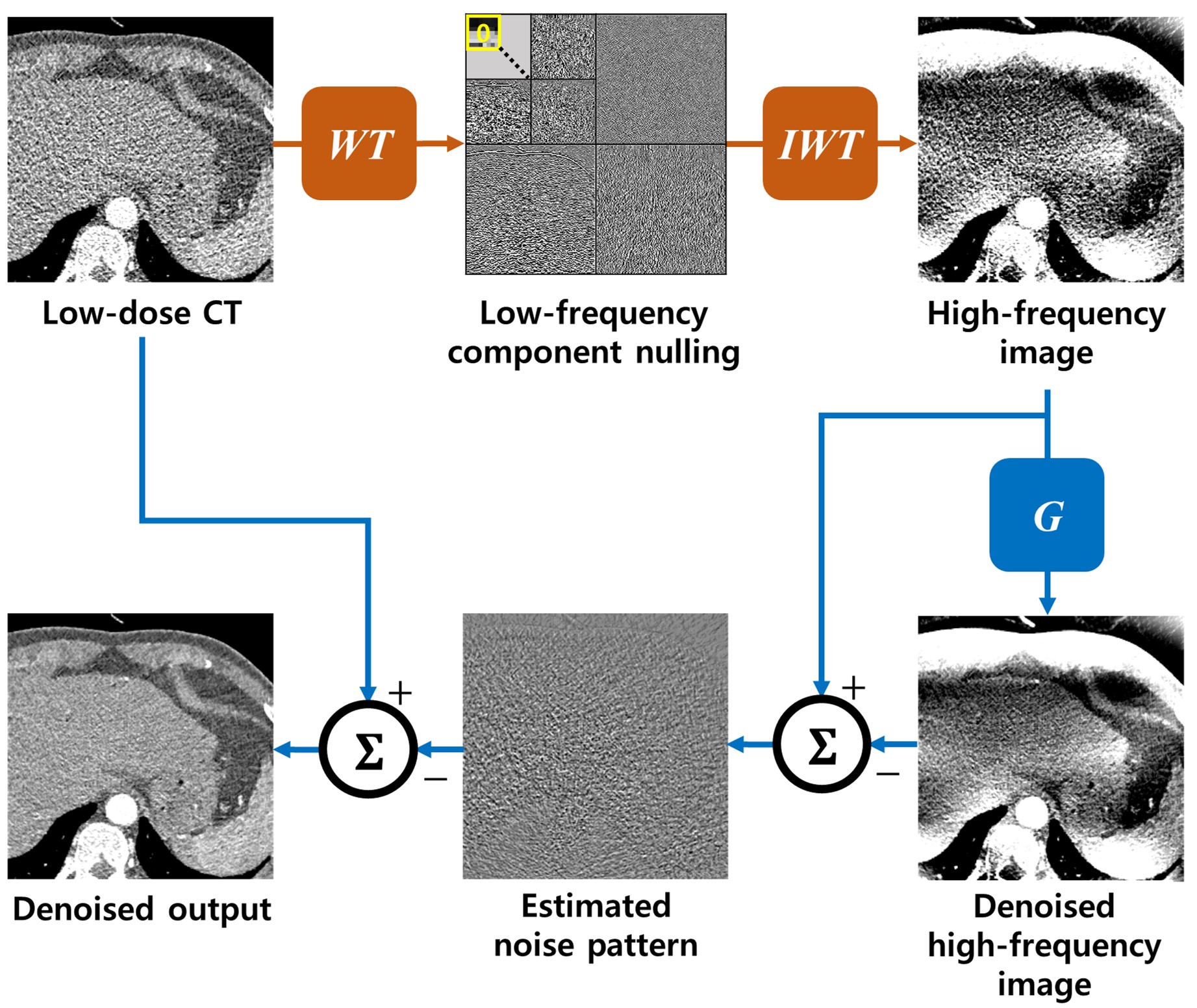}
\caption{Overall framework of denoising method using wavelet transform. WT and IWT denote wavelet transform and inverse wavelet transform, respectively. The last LL component colored in yellow is set to zero to obtain the high-frequency image after IWT. The generator reduces noise in the high-frequency image. Noise pattern is estimated by subtracting the output of the generator from the input. The final denoised CT image is estimated by subtracting the noise pattern from the original low-dose CT image.}
\label{fig_wavelet}
\end{figure}

\subsection{Wavelet High-Frequency Image for Low-Dose CT Denoising}

The method of obtaining the wavelet high-frequency image from the CT image for our method is shown in Fig. \ref{fig_wavelet}. 
The method is similar to the method introduced in the study by Song et al. \cite{song2020unsupervised}.
First, the low-dose CT image is decomposed repeatedly using wavelet transform.
Because the noise in the low-dose CT image has no directionality, 
the wavelet high-frequency image is obtained by inverse wavelet transform of all the detail bands except the last low-low (LL) band.
This process can be easily conducted by zeroing the last LL band.
Our method is different from the previous method \cite{song2020unsupervised} in which only horizontal or vertical bands among wavelet subbands are utilized for network training.

Using our method, non-directional high-frequency noise component can be separated robustly.
The wavelet high-frequency images from the low-dose CT images and the routine-dose CT images are used for unsupervised learning.
After the network removes the noise components of the wavelet high-frequency images, noise pattern is estimated by subtracting the output of the network from the wavelet high-frequency image.
The final output is obtained by subtracting the estimated noise pattern from the low-dose CT image as shown in Fig.~\ref{fig_wavelet}.

\section{Methods}
\subsection{Data Set}
\subsubsection{Multiphase Cardiac CT Scans}

Cardiac multiphase coronary CT (CTA) is a typical case using low-dose CT.
CTA acquires a series of CT images of a heart so that a doctor can observe cardiac motion, which is essential for the diagnosis of several heart diseases such as valve disease.
If the entire series of the CT images were obtained using routine-dose CT, 
an excessive radiation dose would be applied to the patient. 
Conversely, if the entire series were obtained using low-dose CT, 
there is a risk that accurate diagnosis may be impossible. 
Therefore, generally at least one of the CT images is obtained using routine-dose CT, 
and the remaining images are obtained using low-dose CT, 
so that medical staffs can diagnose the heart condition of the patient using the routine-dose CT images with a low noise level and the low-dose CT images with a high noise level.

We used the CTA scan dataset consisting of 50 mitral valve prolapse patients and 50 coronary artery disease patients, which was used in the study by Kang et al. \cite{kang2019cycle}.
The number of CT images in training data is 36,535, and that in test data is 3,530.
Electrocardiography (ECG)-gated cardiac CT scanning was performed for data acquisition using second generation dual-source CT scanner (Somatom Definition Flash, Siemens). 
The tube current of low-dose CT is 20\% for the current of routine-dose CT. 
CT images were reconstructed using a standard cardiac filter (B26f). 
The low-dose CT images are from phase 2 of the CTA dataset and the routine-dose are from phase 8.
Detailed information of the CTA scan protocol is written in the previous reports \cite{koo2014demonstration, yang2015stress}.

\subsubsection{Chest CT Scans}

The low-dose CT image data released in 2020 by the Mayo clinic \cite{clark2013cancer, LDCTdata2020} was used for quantitative analysis of the network performance. 
Of the 100 chest CT data provided, 50 were used.
45 data is as training data, the remaining five were used as test data.
Total number of CT images in the training data is 14,965, and that in the test data is 1,683. 
All the CT scans were taken at routine-dose levels and low-dose levels. 
For the low-dose level, a 10\% dose of the routine-dose was used. 
More information on this data can be found in the technical note published by Chen et al. \cite{chen2015development}.

\subsection{Implementation Details}

The structure of the image generator is U-Net with AdaIN layers as shown in Fig.~\ref{fig_network}, which consists of 4 stages.
At each stage in the encoder part, the width and height of the feature map are halved, 
and the number of channels is doubled.
Strided convolution with a stride of 2 is used instead of pooling layer in the encoder part.
The feature map of the encoder part is concatenated to the feature map of the decoder part in the same stage.
In the decoder part, \(2 \times 2\) upsampling layers are used.
The feature maps are normalized using AdaIN layers with the fixed vectors or with the output vectors from the AdaIN code generator.

\begin{figure}[b!]
\centering
\includegraphics[width=0.9\columnwidth]{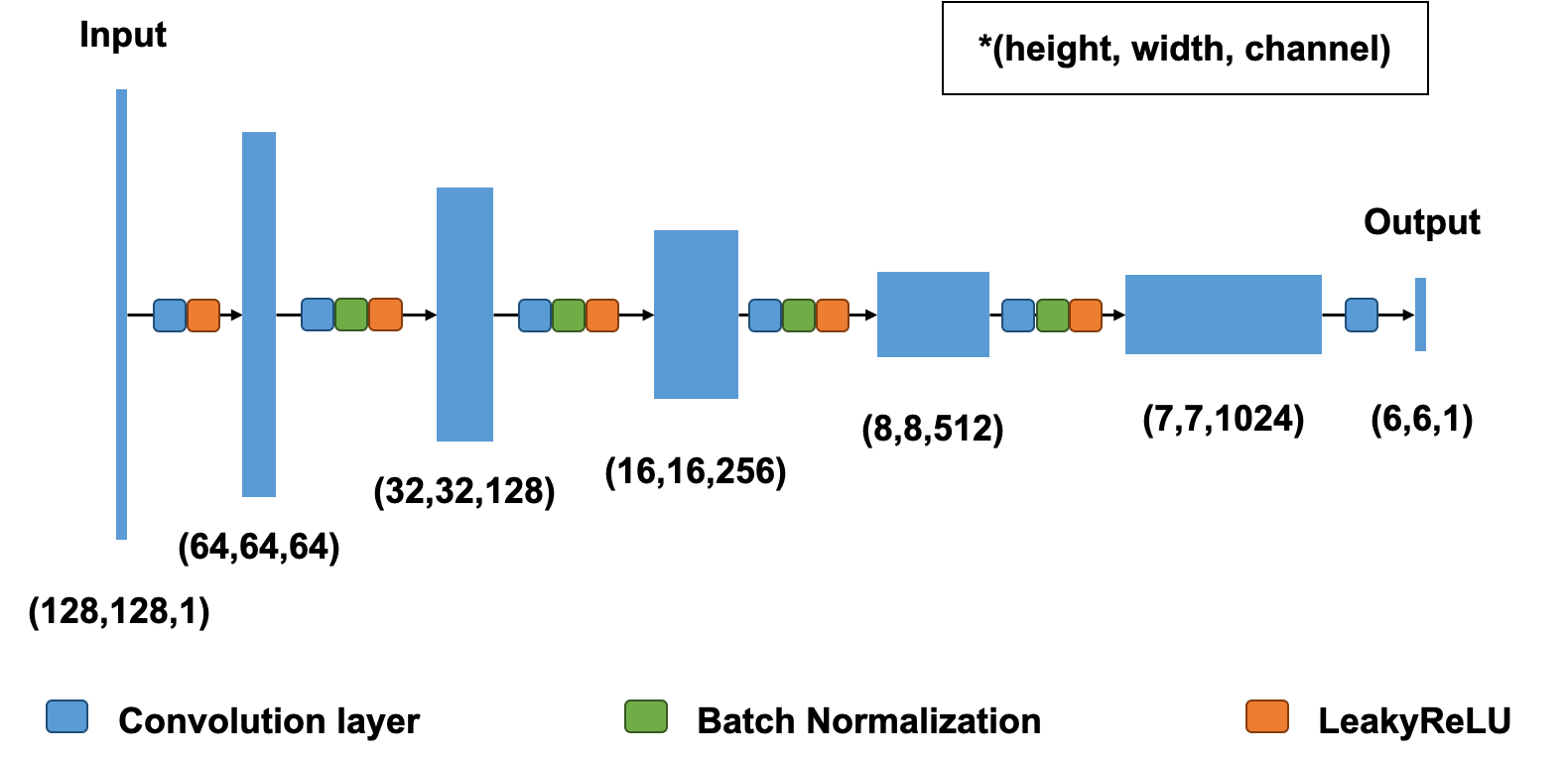}
\caption{Architecture of the discriminator. The generator has a structure of PatchGAN discriminator \cite{isola2017image}. The three numbers below each feature map refers to the height, width, number of channels of the feature map. 
The color coded boxes are the layers forming the network.}
\label{fig_discriminator}
\end{figure}

The discriminator has a similar structure to the discriminator introduced in PatchGAN \cite{isola2017image}.
The architecture of the discriminator is illustrated in Fig. \ref{fig_discriminator}. 
The first four convolution layers use a stride of 2, 
while the rest convolution layers use a stride of 1. 
The first convolution layer gets an input image with one channel 
and generates a feature map with 64 channels.
After that, each time the feature map passes through the convolution layer, 
the number of channels is doubled.
In the last layer, the output is obtained by reducing the number of channels to one.
The discriminator loss \eqref{eq:loss_gan_x} is calculated using the output.

The generators, discriminators and AdaIN code generator were trained by solving \eqref{eq:minmax}.
All the networks were trained using ADAM optimization algorithm \cite{kingma2014adam} with \(\beta_1=0.5\), \(\beta_2=0.999\) and \(\epsilon=1e^{-7}\).
The convolution kernels were initialized by the Glorot uniform initializer \cite{glorot2010understanding}.
The number of mini-batches for training was 1.
All the methods in this study were implemented in Tensorflow v2.2.0 \cite{tensorflow2015-whitepaper} and were trained on NVIDIA GeForce GTX 1080 Ti GPU device.

In the cardiac CT experiment, the loss weights \(\lambda_{cycle}\) and \(\lambda_{identity}\) were set to 10 and 5, respectively.
The number of training epochs was 80. The learning rate was set to \(2e^{-4}\).
In the chest CT experiment, the loss weights \(\lambda_{cycle}\) and \(\lambda_{identity}\) were set to 0.5 and 0.1, respectively. 
The learning rate was set to \(2e^{-3}\), and the training was stopped at 50 epochs.

During training, the randomly cropped patches with size of \(128 \times 128\) were used.
The patches were randomly flopped in horizontally and vertically for data augmentation.
The intensity of CT images was normalized by dividing constant number 1,024 before wavelet transform.
The level of wavelet transform was set to 6 after extensive experiments.
The last LL band was set to zero before inverse wavelet transform to get the wavelet high-frequency image.

\subsection{Comparative Methods}

We compared the performance of the proposed method with those of the various networks.
For the cardiac CT dataset, we compared the proposed method with the cycleGAN from \cite{kang2019cycle}.
The cycleGAN trained two generators with ResNet structure using CT images directly instead of wavelet high-frequency images.
The number of mini-batches for the cycleGAN was 10, and the size of image patches during training was \(56 \times 56\).
Other training settings for the cycleGAN were same to those for the proposed method.

We conducted the experiments using the Chest CT dataset which consists of paired low-dose CT and routine-dose CT image sets for quantitative analysis.
The comparative methods include the supervised method and the cycleGAN from \cite{kang2019cycle}.
The supervised method trained a generator with same structure of the generator of the proposed method, but used batch normalization layers instead of the AdaIN layers.
The generator was trained by minimizing MSE loss between the denoised CT image and the corresponding routine-dose CT image.
The cycleGAN method was trained in the same way of the cycleGAN method for the cardiac CT experiments.

\subsection{Quantitative Analysis}

For quantitative analysis, peak signal-to-noise ratio (PSNR) and the structural similarity index (SSIM) \cite{wang2004image} were calculated. 
PSNR and SSIM values were calculated and averaged for the 5 patients data in the test dataset.
The intensity of the low-dose CT image was normalized to have range [0, 1] by adding the minimum value and dividing the maximum value of the image before calculating PSNR and SSIM values.
The routine-dose CT images and denoised images were normalized in the same way using the values from the low-dose CT image.
The equation for PSNR calculation is as following:
\begin{equation} \label{eq:PSNR}
PSNR(x, y) = 20 \log_{10} \frac{MAX_x}{\|x - y\|_2}
\end{equation}
The SSIM is calculated by
\begin{equation} \label{eq:SSIM}
SSIM(x, y) = \frac{(2\mu_x\mu_y + C_1)(2\sigma_{xy}+C_2)}{({\mu_x}^2+{\mu_y}^2+C_1)({\sigma_x}^2+{\sigma_y}^2+C_2)}
\end{equation}
where \(C_1 = (K_1L)^2\) and \(C_2 = (K_2L)^2\). 
We used \(K_1 = 0.01\), \(K_2 = 0.03\) as in the original paper \cite{wang2004image}.

\section{Experimental results}

\subsection{Cardiac CT Experiment}

\begin{figure}[t!]
\centering
\includegraphics[width=\columnwidth]{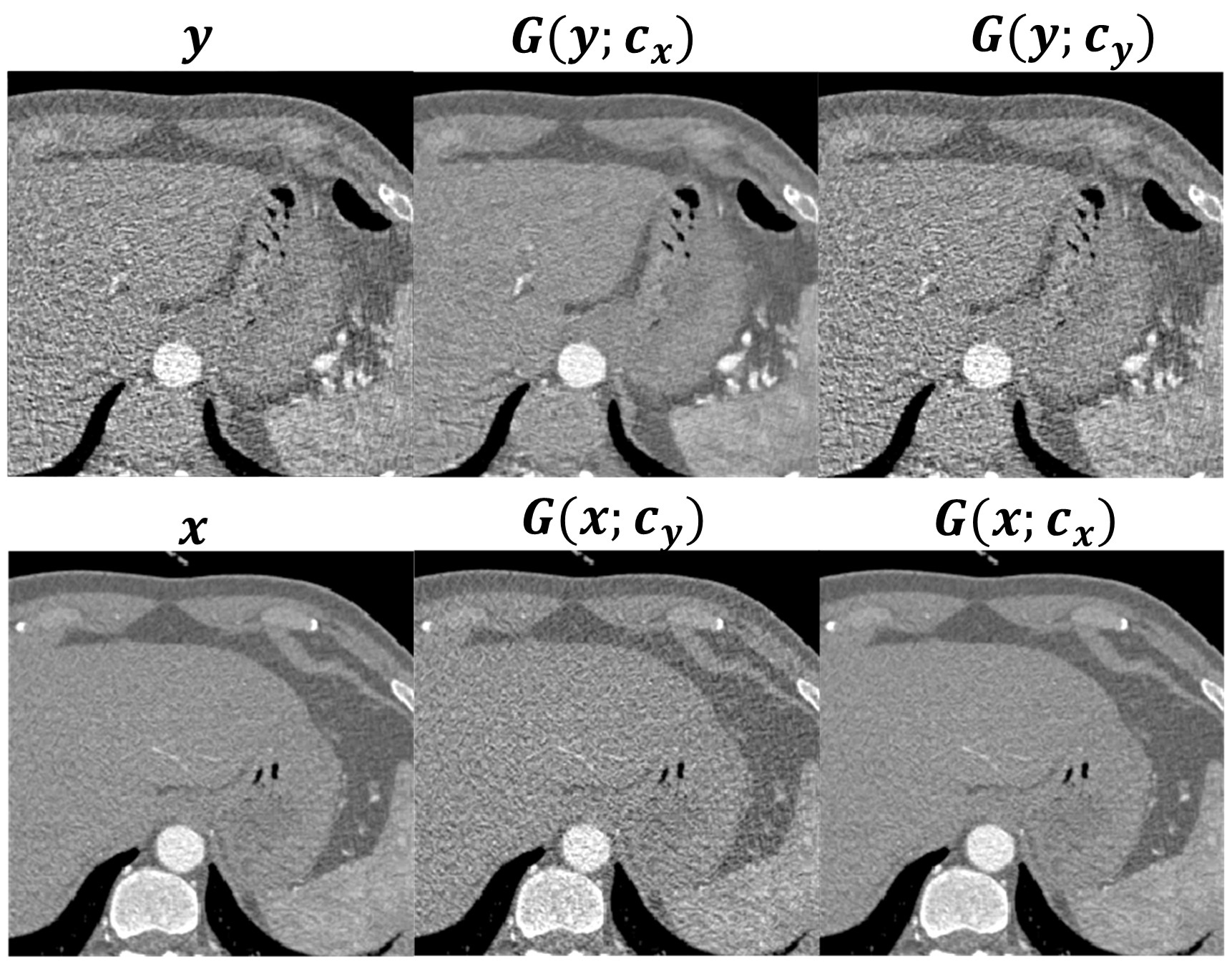}
\caption{Output images of the network with AdaIN. \(y\) is a low-dose CT image, and \(x\) is a routine-dose CT image. Images in the middle column are the output images targeting the other domains. Images in the right column are the output images targeting the same domain. 
The intensity window of CT images is (-500, 500) [HU].}
\label{fig_results_adain}
\end{figure}

\begin{figure}[t!]
\centering
\includegraphics[width=\columnwidth]{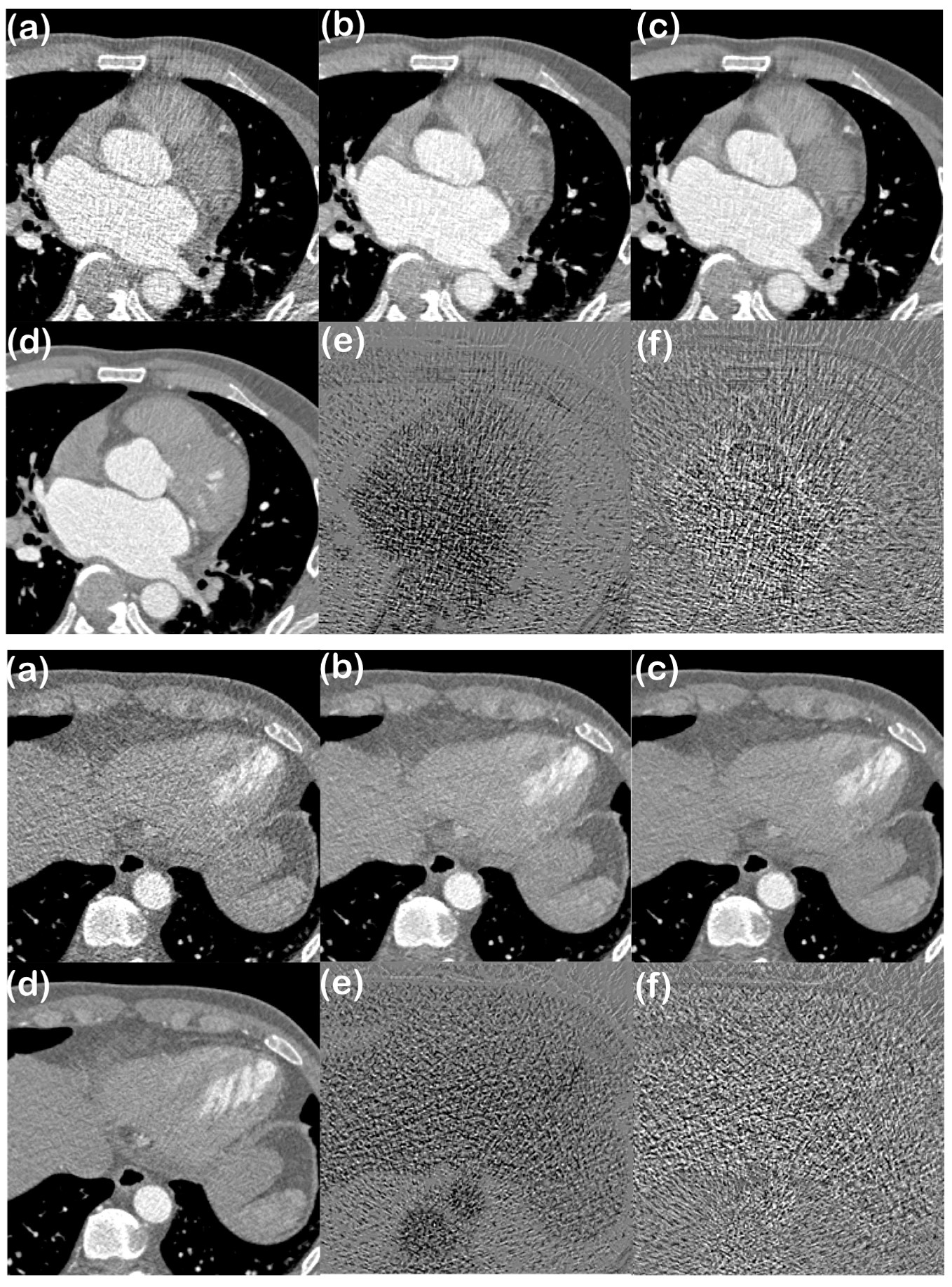}
\caption{Results of the cardiac CT experiment. (a) Input low-dose image, (b) noise reduction results from the cycleGAN method \cite{kang2019cycle}, (c) the proposed method,
and (d) routine-dose images at different cardiac phases. 
(e) Difference image between (b) and (a).  
(f) Difference image between (c) and (a). 
The intensity window of CT image is (-500, 500) [HU] and 
the intensity window of difference image is (-100, 100) [HU]. }
\label{fig_results_multiphase}
\end{figure}

Fig. \ref{fig_results_adain} shows the input images and corresponding output images of the generator with AdaIN layers.
The results for the same input image differ depending on the AdaIN code.
It means that the AdaIN code controls the function of the generator.
The AdaIN code of zero mean and unit variance vectors removes the noise from the input image as shown in the middle column of the first row in Fig. \ref{fig_results_adain}.
The denoised low-dose CT image \(G(y; c_x)\) has a similar noise level to the routine-dose CT image.
In contrast, the AdaIN code \(c_y=F(c)\) generates additional noise on the input image by adjusting the feature map in the network as shown in the image $G(x;c_y)$ 
at the middle column of the second row in Fig. \ref{fig_results_adain}.
On the other hand, thanks to the identity loss,
the images $G(x;c_x)$ and $G(y;c_y)$ do not change much, 
retaining original noise levels as shown in the last column in Fig. \ref{fig_results_adain}.
These results confirm that the network successfully performed the bidirectional image-to-image translation as we intended.

The images of the cardiac CT dataset are arranged in Fig. \ref{fig_results_multiphase}.
Original low-dose CT images and the corresponding routine-dose CT images are displayed in the left column in (a) and (d), respectively.
Note that they are not matched images, since they are taken at different cardiac phases.
Then, the denoised images and the corresponding difference images by the cycleGAN \cite{kang2019cycle} and the propose method are displayed in the second and third column, respectively.
The difference images are obtained by subtracting the denoised images from the low-dose CT images.
The intensity window of the CT images is (-500, 500) 
while the intensity window of the difference images is (-100, 100).
The proposed method removes more noise components than the cycleGAN method, without blurring as shown in the denoised image.
The noises in the difference images of the proposed method are uniformly distributed while those of the cycleGAN are concentrated in the heart.
This confirms that the proposed method is better than the cycleGAN for denoising.

\begin{figure*}[t!]
\centering
\includegraphics[width=0.95\textwidth]{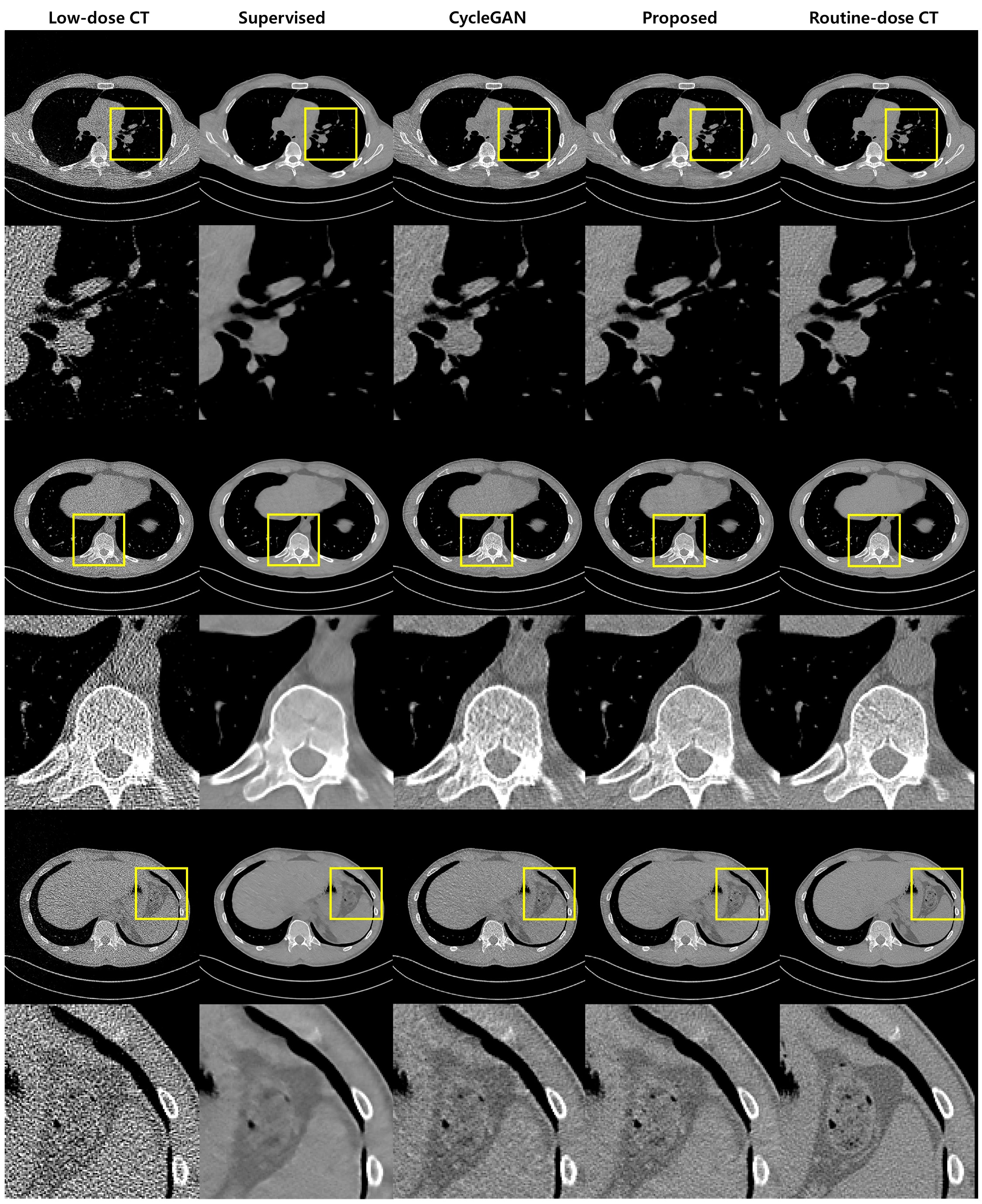}
\caption{Noise reduction results from the Mayo Clinic dataset using the proposed method and the comparative methods. Intensity range of the CT images is (-500, 500) [HU].}
\label{fig_results_mayo}
\end{figure*}

\begin{figure*}[t!]
\centering
\includegraphics[width=0.95\textwidth]{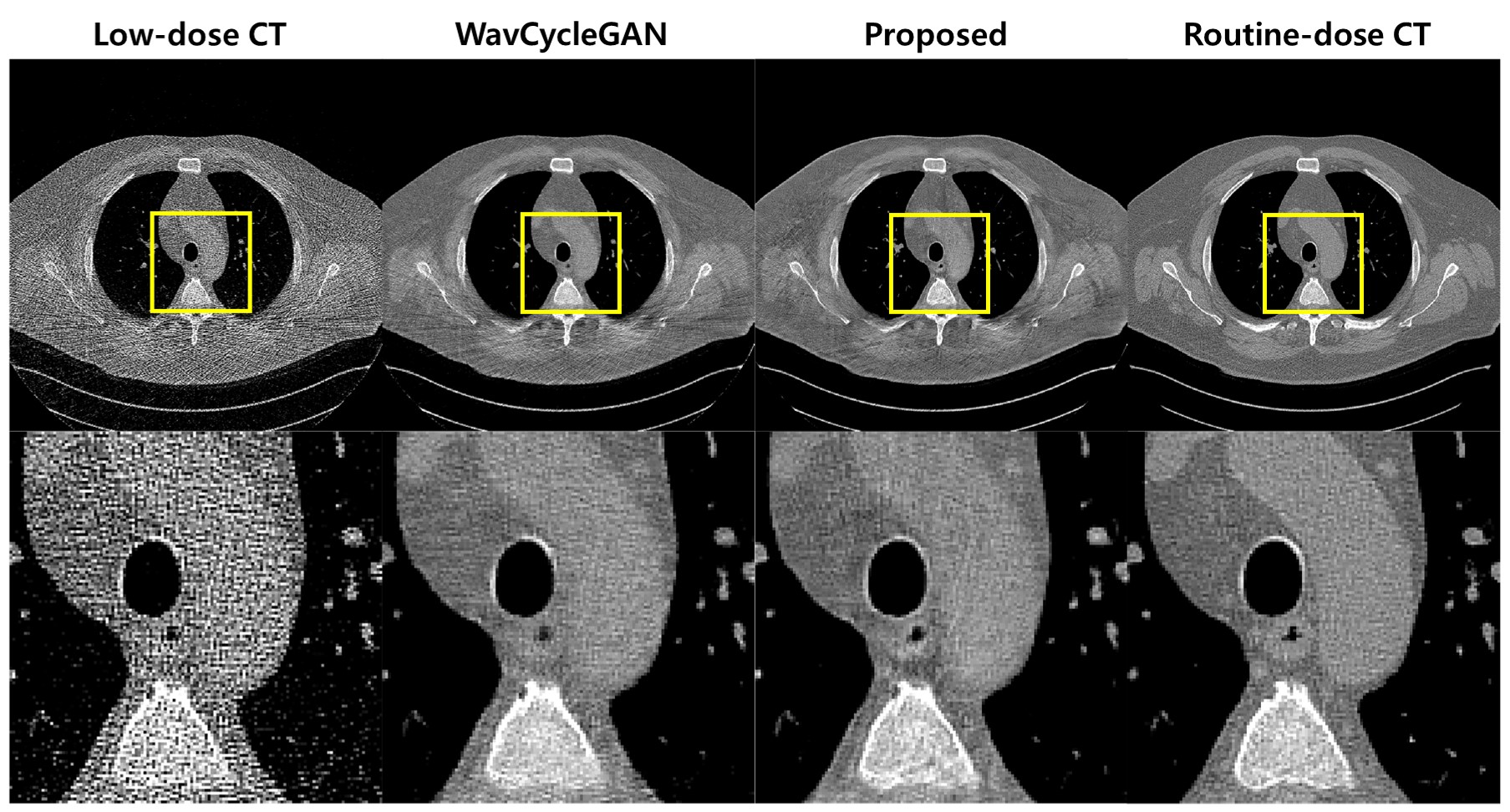}
\caption{Noise reduction results from the Mayo Clinic dataset using the proposed method and the WavCycleGAN. Intensity range of the CT images is (-500, 500) [HU].}
\label{fig:result_wavCycleGAN}
\end{figure*}

\subsection{Chest CT Experiment}

For the chest CT experiment, the images from three comparative methods are arranged in Fig. \ref{fig_results_mayo}.
There are three selective images for each method, 
which show lung, bone and soft tissue from top to bottom.
The image patches of size \(128 \times 128 \) cropped from the selected CT images are displayed below the CT images for detailed observation.
In the results of the supervised method, 
the images are over-smoothed after denoising compared to the routine-dose CT images.
The cycleGAN method still remains the noise on the result images as in the cardiac CT dataset experiment.
The results of the proposed method have a noise level similar to the routine-dose CT images while recovering edges and details without blurring.

\begin{table}[h!]
\centering
\caption{Quantitative comparison of various methods.}
 	\resizebox{0.3\textwidth}{!}{
 \begin{tabular}{c|cc } 
 \hline
 Algorithm & PSNR & SSIM \\
\hline\hline
LDCT input & 22.8585 & 0.3529\\
Supervised & \textbf{33.0848} & \textbf{0.7649} \\
CycleGAN & 30.1970 & 0.6466 \\
Proposed & \textbf{30.8730} & \textbf{0.6605} \\[0.5ex] 
 \hline
 \end{tabular}
 }
\label{table:PSNR_SSIM}
\end{table}

The average PSNR and SSIM values of the various CT images for quantitative comparison are listed in Table \ref{table:PSNR_SSIM}.
Among the methods, the supervised method has the highest PSNR and SSIM values.
However, when you see the results of the supervised method, you can see that the results are blurry and overly smoothed so that the detail information are all removed.
Accordingly, even though the supervised method scores the highest quantitative values,
it is difficult to state that the supervised method is the best method since the images of the supervised method have different texture and noise level to the routine-dose CT images.
Moreover, supervised method is not applicable in general situations where the matched reference data is not available.
Except for the supervised method, the proposed method has the highest PSNR and SSIM values among the unsupervised methods.

\section{Discussion}

To confirm the optimality of the proposed network architecture,
we have performed various ablation studies.
First, we implemented a similar cycleGAN method in wavelet domain using two separate generators (WavCycleGAN).
The structure of the generators used in the WavCycleGAN was same as that of the generator in the supervised method.

Table \ref{table:Parameters} compares the number of parameters in generators for the WavCycleGAN and the proposed method.
As shown in the table, 
the generators for the proposed method only use half the parameters of the WavCycleGAN 
because it uses only one generator and one AdaIN code generator 
which has much smaller number of parameters than the image generator.

\begin{table}[h!]
\centering
\caption{Comparison for number of parameters in generators}
 	\resizebox{0.45\textwidth}{!}{
 \begin{tabular}{c  c | c  c  } 
 \hline
 \multicolumn{2}{c|}{WavCycleGAN} & \multicolumn{2}{c}{Proposed} \\
\hline
 Network & \# of Parameters & Network & \# of Parameters \\
\hline\hline
 \(G_{XY}\) & 5,908,801 & \(G\) & 5,900,865\\
 \(G_{YX}\) & 5,908,801 & \(F\) & 274,560\\
 \hline
 Total  & 11,817,602 & Total & \textbf{6,175,425} \\[0.5ex] 
 \hline
 \end{tabular}
 }
\label{table:Parameters}
\end{table}

The reconstruction results in Fig.~\ref{fig:result_wavCycleGAN} show that 
the noise level of the WavCycleGAN results is similar to that of the routine-dose CT images.
The overall image quality of the WavCycleGAN results is similar to that of the proposed method, but the edges in the WavCycleGAN results are somewhat blurry compared to the proposed method.
Moreover, for the average PSNR and SSIM values, the WavCycleGAN results in PSNR=30.4030 dB and SSIM=0.6279, whereas the proposed method has PSNR=30.8730 dB and SSIM=0.6605.
This shows that the proposed method is comparable and even better than the WavCycleGAN. 
We suspect that the improvement may come from the smaller number of parameters, which makes the learning more stable.

\begin{figure*}[t!]
\centering
\includegraphics[width=0.8\textwidth]{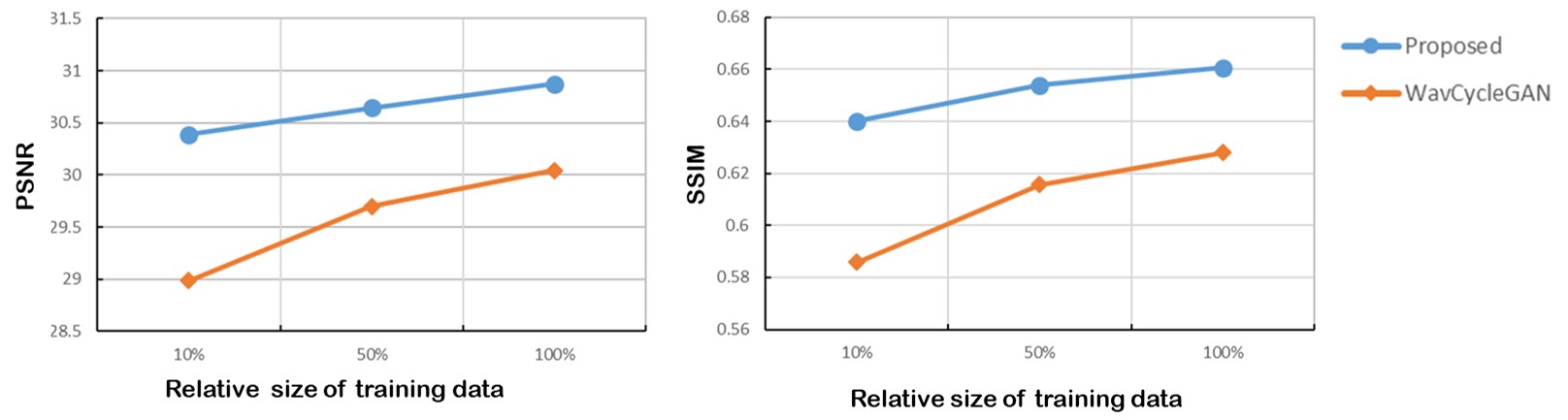}
\caption{PSNR and SSIM plots of the proposed method and the WavCycleGAN according to the relative amount of training dataset.}
\label{fig:reduced}
\end{figure*}

\begin{figure*}[t!]
\centering
\includegraphics[width=0.95\textwidth]{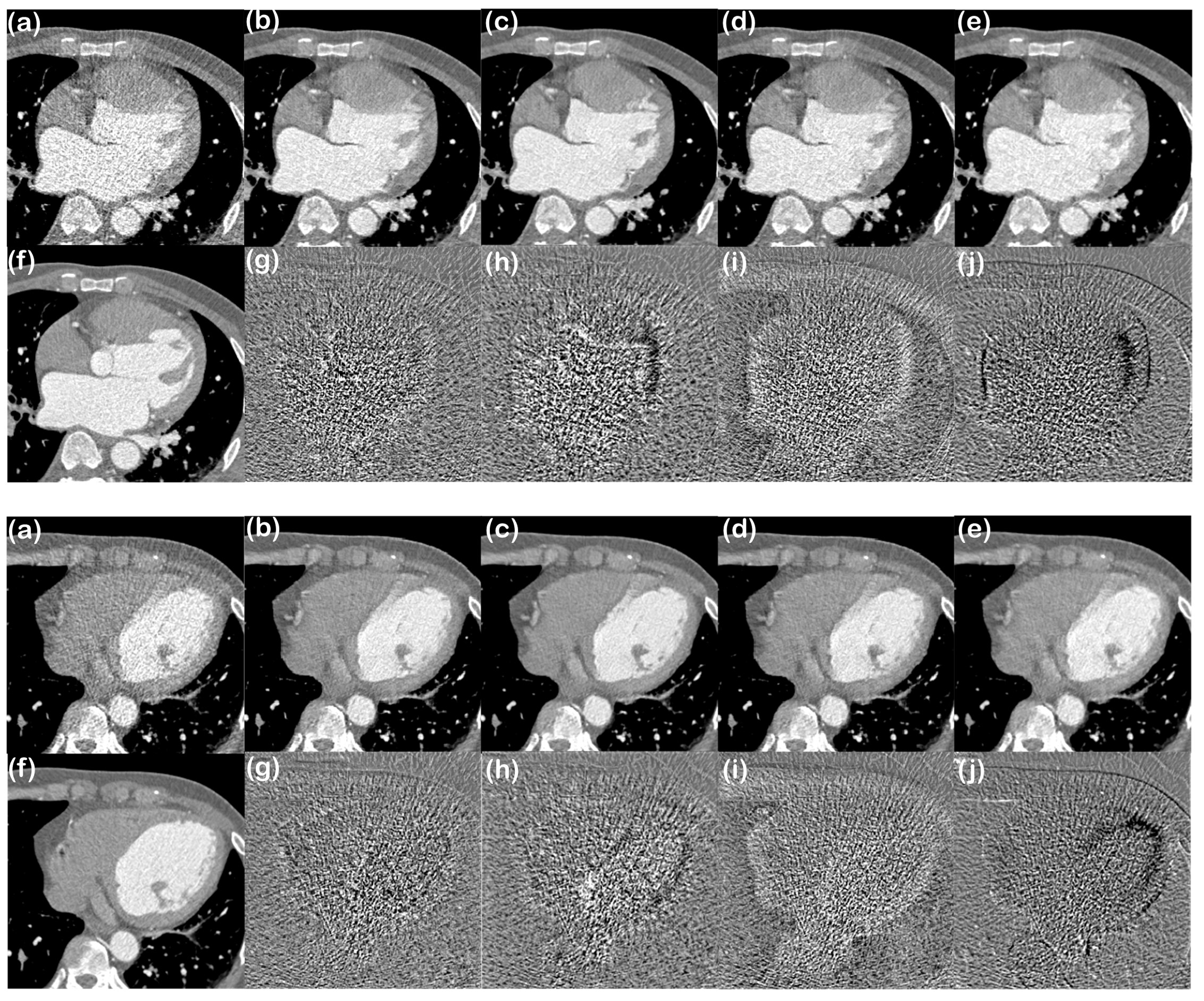}
\caption{Results of ablation studies. (a) Input low-dose CT image, and reconstruction results by (b) the proposed method, (c) the proposed method without wavelet residuals, (d) without identity loss, and (e) without wavelet residuals and identity loss.  (f) Routine-dose CT images at different cardiac phases. (g-j) Difference images between (a) and the results images (b-e). The intensity window of CT image is (-500, 500) [HU] and the intensity window of difference image is (-100, 100) [HU]. }
\label{fig:ablation}
\end{figure*}

To confirm this claim, we conducted an experiment with reduced training dataset.  
In the experiment, we trained the proposed method and the WavCycleGAN using only the 50\% and 10\% of the training dataset, and compared the PSNR and SSIM values.
The graphs in the Fig. \ref{fig:reduced} show that the proposed method results in graceful performance degradation,
whereas the performance of the WavCycleGAN quickly degrades with the reduced training dataset.
The PSNR value of the proposed method decreased from 30.87 dB to 30.64 dB and 30.39 dB 
as the amount of training dataset decreased from 100\% to 50\% and 10\%.
However, the PSNR value of the WavCycleGAN decreased from 30.04 dB to 29.70 dB and 28.98 dB 
under the same condition.
Similarly, the proposed method had SSIM values of 0.6605, 0.6539 and 0.6401 for the 100\%, 50\%, and 10\% of training datasets, while the WavCycleGAN had SSIM values of 0.6279, 0.6156 and 0.5859.

In the second ablation study, 
we used the same single generator with AdaIN layers 
but the processing was done in the image domain rather than wavelet residuals.  Fig.~\ref{fig:ablation} (c) and (h) show that
the image domain approach changes the overall pixel intensity, 
whereas this artifact is not observed in our method in Fig.~\ref{fig:ablation} (b) and (g).

In the last ablation study, we investigated the importance of the identity loss.
Fig.~\ref{fig:ablation} (d) and (i) show the reconstruction results without the identity loss and the corresponding difference images.
In the difference images, you can see the edges between the soft tissue and air, which means that the noise is not uniformly eliminated.
This confirms that the wavelet transform and the identity loss are both important to prevent the generation of artificial features.

\section{Conclusion}
In this paper, we proposed a novel unsupervised learning method using AdaIN-switchable cycleGAN for low-dose CT denoising. 
In the proposed method, the low-dose CT images are first decomposed using wavelet transform so that the neural network retains the low-frequency signal and only learns the noise distributions in the high-frequency regions.
The main contribution of this work is to remove the additional generator in the cycleGAN training process by applying simple AdaIN codes to the generator,
which significantly reduces the number of network weights.
Experimental results confirmed that
although the number of parameters needed for generator training has been reduced by half, the performance of the network has improved
thanks to more stable training.

\section*{Acknowledgment}

The authors would like to thank Dr. Cynthia MaCollough, the Mayo Clinic, and grants EB017095 and EB017185 from the National Institute of Biomedical Imaging and Bioengineering for providing the low-Dose CT and projection dataset.
The authors also thank Dr. Dong Hyun Yang from the University of Ulsan College of Medicine for providing the coronary CT angiography dataset.

\ifCLASSOPTIONcaptionsoff
  \newpage
\fi

\bibliographystyle{IEEEtran}
\bibliography{sample}

\begin{thebibliography}{10}
\providecommand{\url}[1]{#1}
\csname url@samestyle\endcsname
\providecommand{\newblock}{\relax}
\providecommand{\bibinfo}[2]{#2}
\providecommand{\BIBentrySTDinterwordspacing}{\spaceskip=0pt\relax}
\providecommand{\BIBentryALTinterwordstretchfactor}{4}
\providecommand{\BIBentryALTinterwordspacing}{\spaceskip=\fontdimen2\font plus
\BIBentryALTinterwordstretchfactor\fontdimen3\font minus
  \fontdimen4\font\relax}
\providecommand{\BIBforeignlanguage}[2]{{%
\expandafter\ifx\csname l@#1\endcsname\relax
\typeout{** WARNING: IEEEtran.bst: No hyphenation pattern has been}%
\typeout{** loaded for the language `#1'. Using the pattern for}%
\typeout{** the default language instead.}%
\else
\language=\csname l@#1\endcsname
\fi
#2}}
\providecommand{\BIBdecl}{\relax}
\BIBdecl

\bibitem{beister2012iterative}
M.~Beister, D.~Kolditz, and W.~A. Kalender, ``Iterative reconstruction methods
  in x-ray ct,'' \emph{Physica medica}, vol.~28, no.~2, pp. 94--108, 2012.

\bibitem{leipsic2010adaptive}
J.~Leipsic, T.~M. Labounty, B.~Heilbron, J.~K. Min, G.~J. Mancini, F.~Y. Lin,
  C.~Taylor, A.~Dunning, and J.~P. Earls, ``Adaptive statistical iterative
  reconstruction: assessment of image noise and image quality in coronary ct
  angiography,'' \emph{American Journal of Roentgenology}, vol. 195, no.~3, pp.
  649--654, 2010.

\bibitem{renker2011evaluation}
M.~Renker, J.~W. Nance~Jr, U.~J. Schoepf, T.~X. O’Brien, P.~L. Zwerner,
  M.~Meyer, J.~M. Kerl, R.~W. Bauer, C.~Fink, T.~J. Vogl \emph{et~al.},
  ``Evaluation of heavily calcified vessels with coronary ct angiography:
  comparison of iterative and filtered back projection image reconstruction,''
  \emph{Radiology}, vol. 260, no.~2, pp. 390--399, 2011.

\bibitem{krizhevsky2012imagenet}
A.~Krizhevsky, I.~Sutskever, and G.~E. Hinton, ``Imagenet classification with
  deep convolutional neural networks,'' in \emph{Advances in neural information
  processing systems}, 2012, pp. 1097--1105.

\bibitem{ronneberger2015u}
O.~Ronneberger, P.~Fischer, and T.~Brox, ``U-net: Convolutional networks for
  biomedical image segmentation,'' in \emph{International Conference on Medical
  image computing and computer-assisted intervention}.\hskip 1em plus 0.5em
  minus 0.4em\relax Springer, 2015, pp. 234--241.

\bibitem{zhang2016beyond}
K.~Zhang, W.~Zuo, Y.~Chen, D.~Meng, and L.~Zhang, ``Beyond a gaussian denoiser:
  Residual learning of deep {CNN} for image denoising,'' \emph{arXiv preprint
  arXiv:1608.03981}, 2016.

\bibitem{kim2015accurate}
J.~Kim, J.~K. Lee, and K.~M. Lee, ``Accurate image super-resolution using very
  deep convolutional networks,'' \emph{arXiv preprint arXiv:1511.04587}, 2015.

\bibitem{shi2016real}
W.~Shi, J.~Caballero, F.~Husz{\'a}r, J.~Totz, A.~P. Aitken, R.~Bishop,
  D.~Rueckert, and Z.~Wang, ``Real-time single image and video super-resolution
  using an efficient sub-pixel convolutional neural network,'' in
  \emph{Proceedings of the IEEE Conference on Computer Vision and Pattern
  Recognition}, 2016, pp. 1874--1883.

\bibitem{kang2017deep}
E.~Kang, J.~Min, and J.~C. Ye, ``A deep convolutional neural network using
  directional wavelets for low-dose x-ray ct reconstruction,'' \emph{Medical
  physics}, vol.~44, no.~10, pp. e360--e375, 2017.

\bibitem{kang2018deep}
E.~Kang, W.~Chang, J.~Yoo, and J.~C. Ye, ``Deep convolutional framelet denosing
  for low-dose ct via wavelet residual network,'' \emph{IEEE transactions on
  medical imaging}, vol.~37, no.~6, pp. 1358--1369, 2018.

\bibitem{chen2017low}
H.~Chen, Y.~Zhang, M.~K. Kalra, F.~Lin, Y.~Chen, P.~Liao, J.~Zhou, and G.~Wang,
  ``Low-dose ct with a residual encoder-decoder convolutional neural network,''
  \emph{IEEE transactions on medical imaging}, vol.~36, no.~12, pp. 2524--2535,
  2017.

\bibitem{yang2018low}
Q.~Yang, P.~Yan, Y.~Zhang, H.~Yu, Y.~Shi, X.~Mou, M.~K. Kalra, Y.~Zhang,
  L.~Sun, and G.~Wang, ``Low-dose ct image denoising using a generative
  adversarial network with wasserstein distance and perceptual loss,''
  \emph{IEEE transactions on medical imaging}, vol.~37, no.~6, pp. 1348--1357,
  2018.

\bibitem{goodfellow2014generative}
I.~Goodfellow, J.~Pouget-Abadie, M.~Mirza, B.~Xu, D.~Warde-Farley, S.~Ozair,
  A.~Courville, and Y.~Bengio, ``Generative adversarial nets,'' in
  \emph{Advances in neural information processing systems}, 2014, pp.
  2672--2680.

\bibitem{zhu2017unpaired}
J.-Y. Zhu, T.~Park, P.~Isola, and A.~A. Efros, ``Unpaired image-to-image
  translation using cycle-consistent adversarial networks,'' in
  \emph{Proceedings of the IEEE international conference on computer vision},
  2017, pp. 2223--2232.

\bibitem{kang2019cycle}
E.~Kang, H.~J. Koo, D.~H. Yang, J.~B. Seo, and J.~C. Ye, ``Cycle-consistent
  adversarial denoising network for multiphase coronary ct angiography,''
  \emph{Medical physics}, vol.~46, no.~2, pp. 550--562, 2019.

\bibitem{sim2019optimal}
B.~Sim, G.~Oh, J.~Kim, C.~Jung, and J.~C. Ye, ``{Optimal Transport, CycleGAN,
  and penalized LS} for unsupervised learning in inverse problems,''
  \emph{arXiv}, pp. arXiv--1909, 2019.

\bibitem{peyre2019computational}
G.~Peyr{\'e}, M.~Cuturi \emph{et~al.}, ``Computational optimal transport,''
  \emph{Foundations and Trends{\textregistered} in Machine Learning}, vol.~11,
  no. 5-6, pp. 355--607, 2019.

\bibitem{villani2008optimal}
C.~Villani, \emph{Optimal transport: old and new}.\hskip 1em plus 0.5em minus
  0.4em\relax Springer Science \& Business Media, 2008, vol. 338.

\bibitem{huang2017arbitrary}
X.~Huang and S.~Belongie, ``Arbitrary style transfer in real-time with adaptive
  instance normalization,'' in \emph{Proceedings of the IEEE International
  Conference on Computer Vision}, 2017, pp. 1501--1510.

\bibitem{karras2019style}
T.~Karras, S.~Laine, and T.~Aila, ``A style-based generator architecture for
  generative adversarial networks,'' in \emph{Proceedings of the IEEE
  conference on computer vision and pattern recognition}, 2019, pp. 4401--4410.

\bibitem{mroueh2019wasserstein}
Y.~Mroueh, ``Wasserstein style transfer,'' \emph{arXiv preprint
  arXiv:1905.12828}, 2019.

\bibitem{song2020unsupervised}
J.~Song, J.-H. Jeong, D.-S. Park, H.-H. Kim, D.-C. Seo, and J.~C. Ye,
  ``Unsupervised denoising for satellite imagery using wavelet subband
  cyclegan,'' \emph{arXiv preprint arXiv:2002.09847}, 2020.

\bibitem{wolterink2017generative}
J.~M. Wolterink, T.~Leiner, M.~A. Viergever, and I.~I{\v{s}}gum, ``Generative
  adversarial networks for noise reduction in low-dose {CT},'' \emph{IEEE
  transactions on medical imaging}, vol.~36, no.~12, pp. 2536--2545, 2017.

\bibitem{zhou2005nonsubsampled}
J.~Zhou, A.~L. Cunha, and M.~N. Do, ``Nonsubsampled contourlet transform:
  construction and application in enhancement,'' in \emph{IEEE International
  Conference on Image Processing 2005}, vol.~1.\hskip 1em plus 0.5em minus
  0.4em\relax IEEE, 2005, pp. I--469.

\bibitem{ulyanov2016instance}
D.~Ulyanov, A.~Vedaldi, and V.~Lempitsky, ``Instance normalization: The missing
  ingredient for fast stylization,'' \emph{arXiv preprint arXiv:1607.08022},
  2016.

\bibitem{mao2017least}
X.~Mao, Q.~Li, H.~Xie, R.~Y. Lau, Z.~Wang, and S.~Paul~Smolley, ``Least squares
  generative adversarial networks,'' in \emph{Proceedings of the IEEE
  international conference on computer vision}, 2017, pp. 2794--2802.

\bibitem{koo2014demonstration}
H.~J. Koo, D.~H. Yang, S.~Y. Oh, J.-W. Kang, D.-H. Kim, J.-K. Song, J.~W. Lee,
  C.~H. Chung, and T.-H. Lim, ``Demonstration of mitral valve prolapse with ct
  for planning of mitral valve repair,'' \emph{Radiographics}, vol.~34, no.~6,
  pp. 1537--1552, 2014.

\bibitem{yang2015stress}
D.~H. Yang, Y.-H. Kim, J.-H. Roh, J.-W. Kang, D.~Han, J.~Jung, N.~Kim, J.~B.
  Lee, J.-M. Ahn, J.-Y. Lee \emph{et~al.}, ``Stress myocardial perfusion ct in
  patients suspected of having coronary artery disease: visual and quantitative
  analysis—validation by using fractional flow reserve,'' \emph{Radiology},
  vol. 276, no.~3, pp. 715--723, 2015.

\bibitem{clark2013cancer}
K.~Clark, B.~Vendt, K.~Smith, J.~Freymann, J.~Kirby, P.~Koppel, S.~Moore,
  S.~Phillips, D.~Maffitt, M.~Pringle \emph{et~al.}, ``The cancer imaging
  archive (tcia): maintaining and operating a public information repository,''
  \emph{Journal of digital imaging}, vol.~26, no.~6, pp. 1045--1057, 2013.

\bibitem{LDCTdata2020}
C.~McCollough, B.~Chen, D.~Holmes, X.~III, Duan, Z.~Yu, L.~Yu, S.~Leng, and
  J.~Fletcher, ``Low dose ct image and projection data [data set],'' 2020, the
  Cancer Imaging Archive.

\bibitem{chen2015development}
B.~Chen, X.~Duan, Z.~Yu, S.~Leng, L.~Yu, and C.~McCollough, ``Development and
  validation of an open data format for ct projection data,'' \emph{Medical
  physics}, vol.~42, no.~12, pp. 6964--6972, 2015.

\bibitem{isola2017image}
P.~Isola, J.-Y. Zhu, T.~Zhou, and A.~A. Efros, ``Image-to-image translation
  with conditional adversarial networks,'' in \emph{Proceedings of the IEEE
  conference on computer vision and pattern recognition}, 2017, pp. 1125--1134.

\bibitem{kingma2014adam}
D.~P. Kingma and J.~Ba, ``Adam: A method for stochastic optimization,''
  \emph{arXiv preprint arXiv:1412.6980}, 2014.

\bibitem{glorot2010understanding}
X.~Glorot and Y.~Bengio, ``Understanding the difficulty of training deep
  feedforward neural networks,'' in \emph{Proceedings of the thirteenth
  international conference on artificial intelligence and statistics}, 2010,
  pp. 249--256.

\bibitem{tensorflow2015-whitepaper}
M.~Abadi, A.~Agarwal, P.~Barham, E.~Brevdo, Z.~Chen, C.~Citro, G.~S. Corrado,
  A.~Davis, J.~Dean, M.~Devin, S.~Ghemawat, I.~Goodfellow, A.~Harp, G.~Irving,
  M.~Isard, Y.~Jia, R.~Jozefowicz, L.~Kaiser, M.~Kudlur, J.~Levenberg,
  D.~Man\'{e}, R.~Monga, S.~Moore, D.~Murray, C.~Olah, M.~Schuster, J.~Shlens,
  B.~Steiner, I.~Sutskever, K.~Talwar, P.~Tucker, V.~Vanhoucke, V.~Vasudevan,
  F.~Vi\'{e}gas, O.~Vinyals, P.~Warden, M.~Wattenberg, M.~Wicke, Y.~Yu, and
  X.~Zheng, ``{TensorFlow}: Large-scale machine learning on heterogeneous
  systems,'' 2015, software available from tensorflow.org.

\bibitem{wang2004image}
Z.~Wang, A.~C. Bovik, H.~R. Sheikh, and E.~P. Simoncelli, ``Image quality
  assessment: from error visibility to structural similarity,'' \emph{IEEE
  transactions on image processing}, vol.~13, no.~4, pp. 600--612, 2004.

\end{thebibliography}

\end{document}